\documentclass[conference, twoside]{IEEEtran}
\usepackage{graphicx}
\usepackage{amssymb}
\usepackage{calc}
\usepackage{array}
\usepackage{subfigure}
\usepackage{graphicx}
\usepackage{mathtools}
\usepackage{multirow}
\usepackage[usenames,dvipsnames]{color}
\usepackage[table]{xcolor}
\definecolor{LightCyan}{rgb}{0.88,1,1}

\begin{document}
\title{\huge Retrospective Interference Alignment for Two-Cell Uplink MIMO Cellular Networks with Delayed CSIT}

\author{
   \IEEEauthorblockN{Wonjae Shin, Yonghee Han, Jungwoo Lee}\vspace{0.5mm}
    \IEEEauthorblockA{Department of Electrical and Computer Eng.\\
    Seoul National University,
     Seoul 151-744, Korea\\
    Email: \{wonjae.shin, murm0723, junglee\}@snu.ac.kr}
  \and
    \IEEEauthorblockN{Namyoon Lee and Robert W. Heath, Jr.}\vspace{0.5mm}
    \IEEEauthorblockA{Department of Electrical and Computer Eng.\\
    The University of Texas at Austin,
    Austin, TX 78712, USA\\
    Email: \{namyoon.lee, rheath\}@utexas.edu}
}
\maketitle

\begin{abstract}
In this paper, we propose a new retrospective interference alignment for two-cell multiple-input multiple-output (MIMO) interfering multiple access channels (IMAC) with the delayed channel state information at the transmitters (CSIT). It is shown that having delayed CSIT can strictly increase the sum-DoF compared to the case of no CSIT. The key idea is to align multiple interfering signals from adjacent cells onto a small dimensional subspace over time by fully exploiting the previously received signals as side information with outdated CSIT in a distributed manner. Remarkably, we show that the retrospective interference alignment can achieve the optimal sum-DoF in the context of two-cell two-user scenario by providing a new outer bound.
\end{abstract}


\section{Introduction}
\label{sec_intro}
Wireless cellular networks are fundamentally limited by interference between multiple cells sharing the same wireless medium. One solution to manage interference is interference alignment (IA), which aligns the multiple interference signals into smaller subspace. IA was initially introduced by \cite{Maddah-Ali}-\cite{Cadambe}, and has been studied for various scenarios such as the X channel, the interference channel, and cellular networks. However, IA relies on instantaneous and global channel state information at the transmitter (CSIT), which is difficult to achieve especially in the future cellular networks adopting frequency-division duplex (FDD) 3GPP LTE/LTE-A \cite{LTE}. In FDD systems, the channel state has to be measured at the receiver and fed back to the transmitter, incurring feedback delay. When the feedback delay is relatively short over the coherence time, the current CSI can be predicted precisely by exploiting the temporal channel correlation. As the coherence time of channels becomes shorter, the CSI becomes completely stale and may not be useful for channel prediction.

Recently, the impact of delayed CSIT was first explored in the pioneering work \cite{Tse}-\cite{Jafar}. In particular,  retrospective IA is an innovative transmission strategy that exploits previously received signals to create signals of common interest to multiple receivers using completely delayed channel knowledge at transmitters, and hence it is capable of significantly improving sum degrees of freedom (sum-DoF) by broadcasting them to the receivers simultaneously. It turns out that even completely delayed CSIT can be very helpful in terms of the sum-DoF for multiple-input single-output (MISO) broadcast channels. To be specific, it was shown that the sum-DoF of the $K$-user MISO broadcast channel with the delayed version of CSI feedback is given by $\frac{K}{1+\frac{1}{2}+\cdots +\frac{1}{K}}$. This is strictly greater than the 1 sum-DoF which can be achieved in the absence of CSIT \cite{Vaze}-\cite{Huang}. Motivated by \cite{Tse}-\cite{Jafar}, there have been several interesting extensions for the interference channels and X channels \cite{Khandani}-\cite{Varanasi}. The sum-DoF gain was characterized for the $K$-user single-input single-output (SISO) interference channel and $2\times K$ SISO X channel under delayed CSIT assumption in \cite{Khandani}. In \cite{Ghasemi}, new achievability and converse bounds for the sum-DoF of the $(N,M)$ multiple-input multiple-output (MIMO) X channel with $N$ antennas at each transmitter and $M$ antennas at each receiver were characterized with delayed CSIT. In particular, \cite{Avestimehr} established the DoF region and sum-DoF of the MIMO X channel for symmetric and asymmetric antenna configurations, respectively, by developing new converses based on a novel \textit{Rank-Ratio Inequality} \cite{Avestimehr2}. Meanwhile, the DoF region of the general MIMO interference channel with an arbitrary number of antennas at each of the four terminals was completely characterized by providing tight inner and outer bounds under the delayed CSIT assumption \cite{Varanasi}. Subsequently, a variety of CSI feedback assumptions, such as moderately delayed CSIT \cite{Lee} and alternating CSIT \cite{Ravi} between instantaneous CSIT and no CSIT setting, have also been investigated, which provide new insight into the interplay between CSI feedback delay and system performance in terms of sum-DoF gain.

In this paper, we devise a new type of retrospective IA in the two-cell MIMO interfering multiple access channels (IMAC)\footnote{The sum-DoF of MIMO-IMAC with instantaneous and global CSIT are partially known in the specific antenna configurations \cite{Suh}-\cite{Berry}.} with $M$ receive antennas and $K$ users per cell each with $N$ transmit antennas with delayed CSIT, which are referred to as two-cell $(M,N,K)$ MIMO-IMAC. It is shown that having delayed CSIT can strictly increase the sum-DoF compared to the case with no CSIT. The key idea is to align multiple interfering signals from the adjacent cell onto a small dimensional subspace over multiple time slots by fully exploiting the past reception signals as useful side information with outdated CSIT in a distributed fashion. By providing a new outer bound using \textit{Rank-Ratio Inequality}, we show that the retrospective IA can achieve the optimal sum-DoF of $\frac{6}{5}M$ for $K=2$ and $M=N$. Our results provide new insights on how to utilize the completely delayed CSI knowledge by offering sum-DoF gain beyond no CSIT case for cellular networks, especially for uplink scenarios.


Throughout this paper, we use $\mathbf{A}^T$ and $\mathbf{A}^{\dag}$ to indicate the transpose and conjugate transpose of a matrix $\mathbf{A}$, respectively. $\mathbf{0}_{M,N}$ indicates an $M \times N$ matrix consisting of all zeros. In addition, $\mathbb{E}\left[\cdot\right]$ represents the expectation operator.

\begin{figure}[t]
    \centerline{\includegraphics[width=7.7cm]{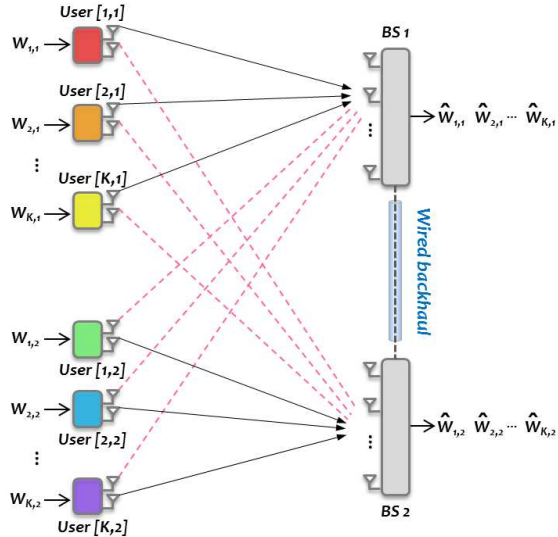}}
    \caption{Two-cell MIMO interfering multiple access channel.}
    \label{ICA}
\end{figure}

\section{System Model}
\label{sec_model}
We consider a system model for the two-cell MIMO-IMAC. Each cell has one base station (BS) and $K$ users
(i.e., mobile stations (MS)) where $K\geq2$. The $k$th user in the $l$th cell is
denoted as user $[k,l]$ for $k \in \{1,2,\cdots,K \}$ and $l \in
\{1,2\}$. Each user is equipped with $N$ antennas, and each
BS is equipped with $M$ antennas where $M,N\geq2$, which will be referred to as the ($M$,$N$,$K$) MIMO-IMAC. To consider a more realistic cellular environment, we shall focus here on $M \geq N$, i.e., the number of antennas at BS is greater than or equal to the number of antennas at user.
As illustrated in Fig. 1, the user $[k,l], k\in\{1,2,\ldots,K\}, l\in\{1,2\}$ intends to send message $W_{k,l}$ to its corresponding BS. This model well captures an uplink cellular network that shares the same frequency band. Due to the simultaneous transmission, the users in cell 1 create co-channel interference to cell 2, and vice versa. We assume that the channels are completely independent across time, and that the delayed equivalent channel knowledge (obtained after applying the received combining vector) can be given at the transmitters (i.e., delayed CSIT at MS) through a noiseless feedback link while the receiver (BS) has global CSI instantaneously (instantaneous global CSIR), i.e., it knows not only the channels associated with itself, but also the channels of the other receiver as well. To be more specific, the delayed equivalent channel knowledge is the local channel coefficients after applying the received combining vector (at BS), which is delayed. Here, the received combining vector should be constructed with the knowledge of global and instantaneous CSI at each BS. For current cellular standards such as 3GPP LTE/LTE-A \cite{LTE}, base stations are connected with wired backhaul such as X2 links, thus making the assumption reasonable.

When a total of $2K$ users simultaneously send their signals at time slot $m$, the received signal $\mathbf{Y}^{[i]}(m)$ at the $i$th BS is
\begin{eqnarray}
\mathbf{Y}^{[i]}(m) = \sum_{l=1}^{2} \sum_{k=1}^{K}
\mathbf{H}_{i}^{[k,l]}(m)
\mathbf{X}^{[k,l]}(m)+
\mathbf{Z}^{[i]}(m),
\end{eqnarray}
where $\mathbf{X}^{[k,l]}(m)\in \mathbb{C}^{N\times 1}$ denotes the signal vector sent by the user $[k,l]$ over the $m$th time slot with an average power constraint, $\mathbb{E}\left[\|\mathbf{X}^{[k,l]}(m)\|^2\right]\leq P$; $\mathbf{H}_{i}^{[k,l]}(m)\in \mathbb{C}^{M \times N}$ represents the channel matrix from the user $[k,l]$ to the BS $i$, the entry of which is independent and identically distributed (i.i.d.) with $\mathcal{CN}(0,1)$; and $\mathbf{Z}^{[i]}(m)\in \mathbb{C}^{M \times 1}$ denotes the additive white Gaussian
noise (AWGN) vector at the $i$th BS with variance $\sigma^{2}$
per entry.

The sum-DoF is defined as the pre-log factor of the achievable sum rate. The
individual DoF achieved by user $[k,l]$ and the sum-DoF are expressed as
\begin{eqnarray}
d^{[k,l]} \triangleq \lim_{\mathrm{SNR} \rightarrow \infty}
\frac{R^{[k,l]}(\mathrm{SNR})}{\mbox{log}(\mathrm{SNR})} ~~\mbox{and}~~
\mathrm{DoF_{sum}} = \sum_{\forall k,l}{d^{[k,l]}},
\end{eqnarray}
 where the SNR is given by $\frac{P}{\sigma^2}$ and $R^{[k,l]}(\mathrm{SNR})$ denotes the achievable rate of $W_{k,l}$ for the average power-constraint $P$.

\section{Retrospective IA using outdated CSIT}
\label{sec_main}
In this section, we introduce a new retrospective IA for two-cell $K$-user MIMO-IMAC using outdated and local CSIT, and characterize the sum-DoF. The proposed transmission strategy consists of 3 phases based on a retrospective IA approach \cite{Tse}-\cite{Jafar}. During phase 1 (phase 2), all users in cell 1 (cell 2) transmit their data streams intended for their corresponding BS. In phase 3, each user sends the linear combinations of past transmissions so that each BS receives a sum of desired message signals and a previously overheard undesired interference signal with the help of the outdated CSIT. Due to page limitations, we only provide the achievability proof for $(M,N,K)=(2,2,2)$ and $(K,2,K)$. For the general achievability, please see the journal version of this paper \cite{Shin2}.

\subsection{Achievable Scheme for $(M,N,K)=(2,2,2)$}
Throughout this example, we will show that $\frac{12}{5}$ sum-DoF can be achieved using completely outdated and local CSIT. To the end, we will show that in total 5 channel uses, all users in cell 1 can successfully send 6 interference-free symbols overall to BS 1, and so do all users in cell 2 to BS 2. Details of the transmission scheme are described below:

\subsubsection{Phase 1} This phase uses 2 time slots. Users [1,1] and [2,1] in cell 1 send the following information symbols:
\begin{eqnarray}
&\mathbf{X}^{[1,1]}(1) =\left[
                \begin{array}{c}
                  a_1 \\
                  a_2 \\
                \end{array}
              \right],&\,\,\,\,\,\
\mathbf{X}^{[1,1]}(2) =\left[
                \begin{array}{c}
                  a_3 \\
                  0 \\
                \end{array}
              \right],
              \\
&\mathbf{X}^{[2,1]}(1) =\left[
                \begin{array}{c}
                  b_1 \\
                  0 \\
                \end{array}
              \right],&\,\,\,\,\,\,
\mathbf{X}^{[2,1]}(2) =\left[
                \begin{array}{c}
                  b_2 \\
                  b_3 \\
                \end{array}
              \right].
\end{eqnarray}
Then, the receivers in this phase have the following resultant input-output relationship:\\
at BS 1,
\begin{eqnarray}
\mathbf{Y}^{[1]}(1)=\mathbf{H}^{[1,1]}_1(1)\mathbf{X}^{[1,1]}(1) + \mathbf{H}^{[2,1]}_1(1)\mathbf{X}^{[2,1]}(1) + \mathbf{Z}^{[1]}(1),\nonumber\\
\mathbf{Y}^{[1]}(2)=\mathbf{H}^{[1,1]}_1(2)\mathbf{X}^{[1,1]}(2) + \mathbf{H}^{[2,1]}_1(2)\mathbf{X}^{[2,1]}(2) + \mathbf{Z}^{[1]}(2),\nonumber
\end{eqnarray}
and at BS 2,
\begin{eqnarray}
\mathbf{Y}^{[2]}(1)=\mathbf{H}^{[1,1]}_2(1)\mathbf{X}^{[1,1]}(1) + \mathbf{H}^{[2,1]}_2(1)\mathbf{X}^{[2,1]}(1) + \mathbf{Z}^{[2]}(1),\nonumber\\
\mathbf{Y}^{[2]}(2)=\mathbf{H}^{[1,1]}_2(2)\mathbf{X}^{[1,1]}(2) + \mathbf{H}^{[2,1]}_2(2)\mathbf{X}^{[2,1]}(2) + \mathbf{Z}^{[2]}(2).\nonumber
\end{eqnarray}
Let us define the short-hand notations:
\begin{eqnarray}
\mathbf{L}^{[1]}_1(a_1,a_2,b_1) &=& \mathbf{H}^{[1,1]}_1(1)\mathbf{X}^{[1,1]}(1) + \mathbf{H}^{[2,1]}_1(1)\mathbf{X}^{[2,1]}(1),\nonumber\\
\mathbf{L}^{[1]}_2(a_3,b_2,b_3) &=& \mathbf{H}^{[1,1]}_1(2)\mathbf{X}^{[1,1]}(2) + \mathbf{H}^{[2,1]}_1(2)\mathbf{X}^{[2,1]}(2),\nonumber\\
\mathbf{L}^{[2]}_1(a_1,a_2,b_1) &=& \mathbf{H}^{[1,1]}_2(1)\mathbf{X}^{[1,1]}(1) + \mathbf{H}^{[2,1]}_2(1)\mathbf{X}^{[2,1]}(1),\nonumber\\
\mathbf{L}^{[2]}_2(a_3,b_2,b_3) &=& \mathbf{H}^{[1,1]}_2(2)\mathbf{X}^{[1,1]}(2) + \mathbf{H}^{[2,1]}_2(2)\mathbf{X}^{[2,1]}(2).\nonumber
\end{eqnarray}
Note that BS 2 saves the overheard equation vectors, $\mathbf{L}^{[2]}_1(a_1,a_2,b_1)$ and $\mathbf{L}^{[2]}_2(a_3,b_2,b_3)$, for use later (in phase 3), although these only carry information intended for the other BS (i.e., BS 1). For simplicity, we drop the noise terms from the received signals, which does not affect the sum-DoF characterization in the high SNR regime.

\subsubsection{Phase 2} The second phase uses 2 time slots, and it is dedicated to the users $[1,2]$ and $[2,2]$ in cell 2. The users send symbols intended to their corresponding BS (i.e., BS 2) as
\begin{eqnarray}
&\mathbf{X}^{[1,2]}(3) =\left[
                \begin{array}{c}
                  c_1 \\
                  c_2 \\
                \end{array}
              \right],&\,\,\,\,\,\
\mathbf{X}^{[1,2]}(4) =\left[
                \begin{array}{c}
                  c_3 \\
                  0 \\
                \end{array}
              \right],
              \\
&\mathbf{X}^{[2,2]}(3) =\left[
                \begin{array}{c}
                  d_1 \\
                  0 \\
                \end{array}
              \right],&\,\,\,\,\,\,
\mathbf{X}^{[2,2]}(4) =\left[
                \begin{array}{c}
                  d_2 \\
                  d_3 \\
                \end{array}
              \right].
\end{eqnarray}
The input-output relationship at the receivers in this phase is described by\\
at BS 1,
\begin{eqnarray}
\mathbf{Y}^{[1]}(3)=\mathbf{H}^{[1,2]}_1(3)\mathbf{X}^{[1,2]}(3) + \mathbf{H}^{[2,2]}_1(3)\mathbf{X}^{[2,2]}(3) + \mathbf{Z}^{[1]}(3),\nonumber\\
\mathbf{Y}^{[1]}(4)=\mathbf{H}^{[1,2]}_1(4)\mathbf{X}^{[1,2]}(4) + \mathbf{H}^{[2,2]}_1(4)\mathbf{X}^{[2,2]}(4) + \mathbf{Z}^{[1]}(4),\nonumber
\end{eqnarray}
and at BS 2,
\begin{eqnarray}
\mathbf{Y}^{[2]}(3)=\mathbf{H}^{[1,2]}_2(3)\mathbf{X}^{[1,2]}(3) + \mathbf{H}^{[2,2]}_2(3)\mathbf{X}^{[2,2]}(3) + \mathbf{Z}^{[2]}(3),\nonumber\\
\mathbf{Y}^{[2]}(4)=\mathbf{H}^{[1,2]}_2(4)\mathbf{X}^{[1,2]}(4) + \mathbf{H}^{[2,2]}_2(4)\mathbf{X}^{[2,2]}(4) + \mathbf{Z}^{[2]}(4).\nonumber
\end{eqnarray}
The short-hand notations are defined as
\begin{eqnarray}
\mathbf{L}^{[1]}_3(c_1,c_2,d_1) &=& \mathbf{H}^{[1,2]}_1(3)\mathbf{X}^{[1,2]}(3) + \mathbf{H}^{[2,2]}_1(3)\mathbf{X}^{[2,2]}(3),\nonumber\\
\mathbf{L}^{[1]}_4(c_3,d_2,d_3) &=& \mathbf{H}^{[1,2]}_1(4)\mathbf{X}^{[1,2]}(4) + \mathbf{H}^{[2,2]}_1(4)\mathbf{X}^{[2,2]}(4),\nonumber\\
\mathbf{L}^{[2]}_3(c_1,c_2,d_1) &=& \mathbf{H}^{[1,2]}_2(3)\mathbf{X}^{[1,2]}(3) + \mathbf{H}^{[2,2]}_2(3)\mathbf{X}^{[2,2]}(3),\nonumber\\
\mathbf{L}^{[2]}_4(c_3,d_2,d_3) &=& \mathbf{H}^{[1,2]}_2(4)\mathbf{X}^{[1,2]}(4) + \mathbf{H}^{[2,2]}_2(4)\mathbf{X}^{[2,2]}(4).\nonumber
\end{eqnarray}
While the overheard equation vectors, $\mathbf{L}^{[1]}_3(c_1,c_2,d_1)$ and $\mathbf{L}^{[1]}_4(c_3,d_2,d_3)$, are not really desired information for BS 1, BS 1 saves the overheard equation vectors for future usage (in phase 3) as side information.

The important observation here is that BS 1 already has two independent linear equations with the three variables $a_1$, $a_2$ and $b_1$,\footnote{Since the receiver has $M$ antennas, so that each equation vectors $\mathbf{L}^{[i]}_j(\cdot)$ contains $M$ independent linear equations.} and it requires one more equation to resolve the desired symbols. If BS 1 somehow has any linear combination of equation vectors, $\mathbf{L}^{[2]}_1(a_1,a_2,b_1)$, overheard by BS 2 in phase 1, then it will have enough equations to solve for its intended symbols. In addition, BS 1 also needs one extra equation to be able to resolve its remaining desired information symbols $a_3$, $b_2$, and $b_3$, thus a linear combination of $\mathbf{L}^{[2]}_2(a_3,b_2,b_3)$, overheard by BS 2 during phase 1, can be very useful for BS 1. BS 2 can cancel out $b_1$ and $a_3$ symbols from its overheard received signal vectors at time slots 1 and 2, respectively, to apply the retrospective IA in phase 3 as follows:
\begin{eqnarray}
\widehat{L}^{[2]}_1(a_1,a_2)&=&\mathbf{u}^{[1]\dag}_1\mathbf{L}^{[2]}_1(a_1,a_2,b_1),\\
\widehat{L}^{[2]}_2(b_2,b_3)&=&\mathbf{u}^{[1]\dag}_2\mathbf{L}^{[2]}_2(a_3,b_2,b_3),
\end{eqnarray}
where $\mathbf{u}^{[i]}_j$ is the $M\times1$ combining vector for the two overheard equations at time slot $j$ during phase $i$. The new linear equation $\widehat{L}^{[i]}_j(\cdot)$ has to solely involve symbols transmitted by one transmitter so that these can be locally generated at one transmitter with the delayed and local CSIT. To the end, the linear combiner, $\mathbf{u}^{[1]}_j$, need to satisfy the following condition:
\begin{eqnarray}
\mathbf{u}^{[1]\dag}_j\mathbf{H}^{[\bar{j},1]}_2(j)\mathbf{X}^{[\bar{j},1]}(j) = 0,\,\,\,\forall j,\bar{j}\in\{1,2\},j\neq\bar{j}.
\end{eqnarray}

 Similarly, BS 2 needs to have a linear combination of the overheard equation vectors, $\mathbf{L}^{[1]}_3(c_1,c_2,d_1)$, as well as one of $\mathbf{L}^{[1]}_4(c_3,d_2,d_3)$ from BS 1 in phase 2 so that is has enough equations to detect all desired symbols, $c_i$ and $d_i$, $i\in\{1,2,3\}$. To purify the two overheard equation vectors, we form new equations as follows:
 \begin{eqnarray}
\widehat{L}^{[1]}_3(c_1,c_2)&=&\mathbf{u}^{[2]\dag}_3\mathbf{L}^{[1]}_3(c_1,c_2,d_1),\\
\widehat{L}^{[1]}_4(d_2,d_3)&=&\mathbf{u}^{[2]\dag}_4\mathbf{L}^{[1]}_4(c_3,d_2,d_3),
\end{eqnarray}
where $
\mathbf{u}^{[2]\dag}_{j+2}\mathbf{H}^{[\bar{j},2]}_1(j+2)\mathbf{X}^{[\bar{j},2]}(j+2) = 0,\,\,j,\bar{j}\in\{1,2\},j\neq\bar{j}$.

Therefore, the main goal of phase 3 is to swap these four linear equations (i.e., $\widehat{L}^{[2]}_1(a_1,a_2)$ and $\widehat{L}^{[2]}_2(b_2,b_3)$ to BS 1, and $\widehat{L}^{[1]}_3(c_1,c_2)$ and $\widehat{L}^{[1]}_4(d_2,d_3)$ to BS 2) through the distributed transmitter where each transmitter has access only to the local channel coefficients by a unit delay.

\begin{figure*}[t]
{\footnotesize \begin{eqnarray}
{\label{Eq1}
\left[
     \begin{array}{c}
       \mathbf{Y}^{[1]}(1) \\
       \mathbf{Y}^{[1]}(2) \\
       \mathbf{Y}^{[1]}(3) \\
       \mathbf{Y}^{[1]}(4) \\
       \mathbf{Y}^{[1]}(5) \\
     \end{array}
   \right]
}
&=&
\left[
     \begin{array}{ccc}
       \mathbf{h}^{[1,1]}_{1,1}(1) & \mathbf{h}^{[1,1]}_{1,2}(1) & \mathbf{0}_{M,1}\\
       \mathbf{0}_{M,1} & \mathbf{0}_{M,1} & \mathbf{h}^{[1,1]}_{1,1}(2) \\
       \mathbf{0}_{M,1} & \mathbf{0}_{M,1} & \mathbf{0}_{M,1} \\
       \mathbf{0}_{M,1} & \mathbf{0}_{M,1} & \mathbf{0}_{M,1} \\
       \mathbf{\Upsilon}^{[1]}_1\mathbf{h}^{[1,1]}_{2,1}(1) & \mathbf{\Upsilon}^{[1]}_1\mathbf{h}^{[1,1]}_{2,2}(1) & \mathbf{0}_{M,1}
     \end{array}
\right]
\left[
     \begin{array}{c}
       a_1 \\
       a_2 \\
       a_3 \\
      \end{array}
\right]
+
\left[
     \begin{array}{ccc}
       \mathbf{h}^{[2,1]}_{1,1}(1) & \mathbf{0}_{M,1} & \mathbf{0}_{M,1}\\
       \mathbf{0}_{M,1} & \mathbf{h}^{[2,1]}_{1,1}(2) & \mathbf{h}^{[2,1]}_{1,2}(2)\\
       \mathbf{0}_{M,1} & \mathbf{0}_{M,1} & \mathbf{0}_{M,1} \\
       \mathbf{0}_{M,1} & \mathbf{0}_{M,1} & \mathbf{0}_{M,1} \\
       \mathbf{0}_{M,1} & \mathbf{\Upsilon}^{[1]}_2\mathbf{h}^{[2,1]}_{2,1}(2) & \mathbf{\Upsilon}^{[1]}_2\mathbf{h}^{[2,1]}_{2,2}(2)
     \end{array}
\right]
\left[
     \begin{array}{c}
       b_1 \\
       b_2 \\
       b_3 \\
      \end{array}
\right]\\
&+&
\left[
     \begin{array}{ccc}
       \mathbf{0}_{M,1} & \mathbf{0}_{M,1} & \mathbf{0}_{M,1}\\
       \mathbf{0}_{M,1} & \mathbf{0}_{M,1} & \mathbf{0}_{M,1}\\
       \mathbf{h}^{[1,2]}_{1,1}(3) & \mathbf{h}^{[1,2]}_{1,2}(3) & \mathbf{0}_{M,1} \\
       \mathbf{0}_{M,1} & \mathbf{0}_{M,1} & \mathbf{h}^{[1,2]}_{1,1}(4) \\
       \mathbf{\Upsilon}^{[2]}_3\mathbf{h}^{[1,2]}_{1,1}(3) & \mathbf{\Upsilon}^{[2]}_3\mathbf{h}^{[1,2]}_{1,2}(3) & \mathbf{0}_{M,1}
     \end{array}
\right]
\left[
     \begin{array}{c}
       c_1 \\
       c_2 \\
       c_3 \\
      \end{array}
\right]
+
\left[
     \begin{array}{ccc}
       \mathbf{0}_{M,1} & \mathbf{0}_{M,1} & \mathbf{0}_{M,1}\\
       \mathbf{0}_{M,1} & \mathbf{0}_{M,1} & \mathbf{0}_{M,1}\\
       \mathbf{h}^{[2,2]}_{1,1}(1) & \mathbf{0}_{M,1} & \mathbf{0}_{M,1} \\
       \mathbf{0}_{M,1} & \mathbf{h}^{[2,2]}_{1,1}(4) & \mathbf{h}^{[2,2]}_{1,2}(4) \\
       \mathbf{0}_{M,1} & \mathbf{\Upsilon}^{[2]}_4\mathbf{h}^{[2,2]}_{1,1}(4) & \mathbf{\Upsilon}^{[2]}_4\mathbf{h}^{[2,2]}_{1,2}(4)
     \end{array}
\right]
\left[
     \begin{array}{c}
       d_1 \\
       d_2 \\
       d_3 \\
      \end{array}
\right]
+
\left[
     \begin{array}{c}
       \mathbf{Z}^{[1]}(1) \\
       \mathbf{Z}^{[1]}(2) \\
       \mathbf{Z}^{[1]}(3) \\
       \mathbf{Z}^{[1]}(4) \\
       \mathbf{Z}^{[1]}(5) \\
      \end{array}
\right]\nonumber
\end{eqnarray}
}

{\footnotesize \begin{eqnarray}
{\label{Eq2}
\left[
     \begin{array}{c}
       \mathbf{Y}^{[1]}(1) \\
       \mathbf{Y}^{[1]}(2) \\
       \mathbf{Y}^{[1]}(5) - \mathbf{\Upsilon}^{[2]}_3 \mathbf{Y}^{[1]}(3) - \mathbf{\Upsilon}^{[2]}_4\mathbf{Y}^{[1]}(4)\\
     \end{array}
   \right]
}
&=&
\underbrace{
\left[
     \begin{array}{ccc}
       \mathbf{h}^{[1,1]}_{1,1}(1) & \mathbf{h}^{[1,1]}_{1,2}(1) & \mathbf{0}_{M,1}\\
       \mathbf{0}_{M,1} & \mathbf{0}_{M,1} & \mathbf{h}^{[1,1]}_{1,1}(2) \\
       \mathbf{\Upsilon}^{[1]}_1\mathbf{h}^{[1,1]}_{2,1}(1) & \mathbf{\Upsilon}^{[1]}_1\mathbf{h}^{[1,1]}_{2,2}(1) & \mathbf{0}_{M,1}
     \end{array}
\right]}_{\mathrm{rank}=3}
\left[
     \begin{array}{c}
       a_1 \\
       a_2 \\
       a_3 \\
      \end{array}
\right]
\\&+&
\underbrace{
\left[
     \begin{array}{ccc}
       \mathbf{h}^{[2,1]}_{1,1}(1) & \mathbf{0}_{M,1} & \mathbf{0}_{M,1}\\
       \mathbf{0}_{M,1} & \mathbf{h}^{[2,1]}_{1,1}(2) & \mathbf{h}^{[2,1]}_{1,2}(2)\\
       \mathbf{0}_{M,1} & \mathbf{\Upsilon}^{[1]}_2\mathbf{h}^{[2,1]}_{2,1}(2) & \mathbf{\Upsilon}^{[1]}_2\mathbf{h}^{[2,1]}_{2,2}(2)
     \end{array}
\right]}_{\mathrm{rank}=3}
\left[
     \begin{array}{c}
       b_1 \\
       b_2 \\
       b_3 \\
      \end{array}
\right]\nonumber
+
\left[
     \begin{array}{c}
       \mathbf{Z}^{[1]}(1) \\
       \mathbf{Z}^{[1]}(2) \\
       \mathbf{Z}^{[1]}(5) -\mathbf{\Upsilon}^{[2]}_3\mathbf{Z}^{[1]}(3) - \mathbf{\Upsilon}^{[2]}_4 \mathbf{Z}^{[1]}(4)\\
      \end{array}
\right]\nonumber
\end{eqnarray}
}
\hrulefill
\end{figure*}

\subsubsection{Phase 3} The phase operates in one channel use. We note that at this time, each transmitter is aware of the local channel state information in the past time slots. By taking advantage of the delayed and local CSIT, each transmitter can reconstruct the additional linear equation generated at the end of phase 2 based on the overheard equation vectors by the interfering BS, and it sends the information symbols as follows:
\begin{eqnarray}
\small\mathbf{X}^{[1,1]}(5) =\left[
                \begin{array}{c}
                  \widehat{L}^{[2]}_1(a_1, a_2) \\
                  0 \\
                \end{array}
              \right],
\mathbf{X}^{[2,1]}(5) =\left[
                \begin{array}{c}
                  \widehat{L}^{[2]}_2(b_2, b_3) \\
                  0 \\
                \end{array}
              \right],
              \\
\small\mathbf{X}^{[1,2]}(5) =\left[
                \begin{array}{c}
                  \widehat{L}^{[1]}_3(c_1, c_2) \\
                  0 \\
                \end{array}
              \right],
\mathbf{X}^{[2,2]}(5) =\left[
                \begin{array}{c}
                  \widehat{L}^{[1]}_4(d_2, d_3) \\
                  0 \\
                \end{array}
              \right].
\end{eqnarray}
At receivers, we have
{\begin{eqnarray}
\mathbf{Y}^{[1]}(5)=\mathbf{h}^{[1,1]}_{1,1}(5)
\widehat{L}^{[2]}_1(a_1, a_2)
+ \mathbf{h}^{[2,1]}_{1,1}(5)
\widehat{L}^{[2]}_2(b_2, b_3) \hspace{5.5mm} \\
+\mathbf{h}^{[1,2]}_{1,1}(5)
\widehat{L}^{[1]}_3(c_1, c_2)
+ \mathbf{h}^{[2,2]}_{1,1}(5)
\widehat{L}^{[1]}_4(d_2, d_3) +\mathbf{Z}^{[1]}(5)\nonumber\hspace{-7.5mm}\\
\mathbf{Y}^{[2]}(5)=\mathbf{h}^{[1,1]}_{2,1}(5)
\widehat{L}^{[2]}_1(a_1, a_2)
+ \mathbf{h}^{[2,1]}_{2,1}(5)
\widehat{L}^{[2]}_2(b_2, b_3) \hspace{5.5mm} \\
+\mathbf{h}^{[1,2]}_{2,1}(5)
\widehat{L}^{[1]}_3(c_1, c_2)
+ \mathbf{h}^{[2,2]}_{2,1}(5)
\widehat{L}^{[1]}_4(d_2, d_3)
+\mathbf{Z}^{[2]}(5) \nonumber\hspace{-7.5mm}
\end{eqnarray}
where $\mathbf{h}^{[k,l]}_{i,j}(m)$ is the $j$th column of $\mathbf{H}^{[k,l]}_{i}(m)$.

Consequently, we have completely designed all transmit signals over the 5 time slots in the network. Putting everything together, the received signals of BS 1 during phase 1, 2, and 3 are shown in equation (\ref{Eq1}) on the top of the page, where $\mathbf{\Upsilon}^{[i]}_j$ denotes $\mathbf{h}^{[\mathrm{mod}(j-1,2)+1,\bar{i}]}_{1,1}(5)\mathbf{u}^{[i]\dag}_j$, $\forall i,\bar{i}\in\{1,2\},i\neq\bar{i}$. Note that it is easy to see that at BS 1 the two interference streams $c_1$ and $c_2$ are aligned along with $\left[\mathbf{0}_{1,M}, \mathbf{0}_{1,M}, \big(\mathbf{\Upsilon}^{[2]}_3\mathbf{h}^{[1,2]}_{1,1}(3)\big)^T, \mathbf{0}_{1,M}, \big(\mathbf{\Upsilon}^{[2]}_3\mathbf{h}^{[1,2]}_{1,1}(3)\big)^T\right]^{T}$ after multiplying $\mathbf{\Upsilon}^{[2]}_3$ with $\mathbf{Y}^{[1]}(3)$. Similarly, two symbols $d_2$ and $d_3$ are also aligned by applying the combiner $\mathbf{\Upsilon}^{[2]}_4$ at $\mathbf{Y}^{[1]}(4)$. Using the important observation, the inter-cell interference at BS 1 can be completely eliminated by subtracting both the third and the fourth received signals after multiplying $\mathbf{\Upsilon}^{[2]}_3$ and $\mathbf{\Upsilon}^{[2]}_4$, respectively (including \textit{purifying process} according to (10)-(11)) from the fifth one as shown in equation (\ref{Eq2}). Also, it is straightforward to verify that two channel matrices in (\ref{Eq2}) are linearly independent with probability one due to the fact that all channel values are generic, so that the receiver can successfully decode 6 data symbols $a_i$ and $b_i$, $i\in\{1,2,3\}$ intended for BS 1 by properly removing its corresponding inter-user interference. BS 2 can have the same results as BS 1 since the system is symmetric.
In summary, $\frac{12}{5}$ sum-DoF can be achieved in this example.

\begin{figure*}[t]
    \centerline{\includegraphics[width=18.0cm]{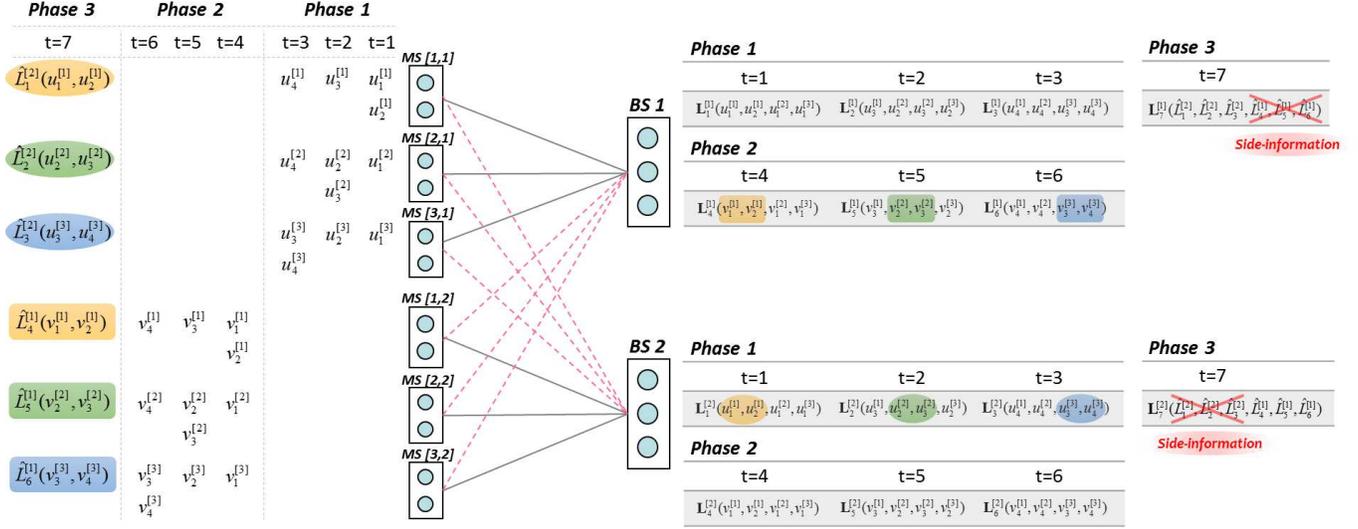}}
    \caption{Achieving $24/7$ sum-DoF for $(3,2,3)$ MIMO-IMAC with delayed and local CSIT.}
    \label{ICA}
\end{figure*}

\vspace{0.0mm}
\textbf{\textit{Remark 1} [An Extension to $(M,M,2)$ MIMO-IMAC]:} Based on the retrospective IA developed in this subsection, one can easily prove that the $(M,M,2)$ MIMO-IMAC can achieve $\frac{6}{5}M$ sum-DoF for an arbitrary $M\geq 2$. 

\subsection{Achievable Scheme for (M,N,K)=(K,2,K)}
Now, we focus on the $(K,2,K)$ MIMO-IMAC. For this case, we will show the achievability of $\frac{2K(K+1)}{2K+1}$, i.e., we show that a total of $2K(K+1)$ information symbols can be transmitted in $2K+1$ channel uses. To show the achievability, the proposed retrospective IA is performed in three phases as depicted in Fig. 2 for the case $K=3$. Phase 1 (Phase 2) is dedicated to users in cell 1 (cell 2) and of duration $K$, where at the time slot $m$, the $[j,i]$ user sends an information as follows:
\begin{eqnarray}\small
\bullet&\mathrm{Phase}\,\,1:& \mathbf{X}^{[j,1]}(m) = \left\{
                            \begin{array}{cc}
                              \left[
                \begin{array}{c}
u^{[j]}_m\\0
                \end{array}
              \right]& \mathrm{for}\,\,\, m<j,\vspace{1mm} \\
                              \left[
                \begin{array}{c}
u^{[j]}_{m}\\u^{[i]}_{m+1}
                \end{array}
              \right]& \mathrm{for}\,\,\, m=j,\vspace{1mm}\nonumber \\
                            \left[
                \begin{array}{c}
u^{[j]}_{m+1}\\0
                \end{array}
              \right]& \mathrm{for}\,\,\, m>j, \\
                            \end{array}
                          \right.  \\
                          &&\mathrm{where} \,\, 1\leq m \leq K, \nonumber
\end{eqnarray}
\begin{eqnarray}\small
\bullet&\mathrm{Phase}\,\,2:& \mathbf{X}^{[j,2]}(m) = \left\{
                            \begin{array}{cc}
                              \left[
                \begin{array}{c}
v^{[j]}_{m'}\\0
                \end{array}
              \right]& \mathrm{for}\,\,\, m'<j,\vspace{1mm} \\
                              \left[
                \begin{array}{c}
v^{[j]}_{m'}\\u^{[i]}_{m'+1}
                \end{array}
              \right]& \mathrm{for}\,\,\, m'=j,\vspace{1mm} \nonumber\\
                            \left[
                \begin{array}{c}
v^{[j]}_{m'+1}\\0
                \end{array}
              \right]& \mathrm{for}\,\,\, m'>j. \\
                            \end{array}
                          \right.\\
                          &&\mathrm{where} \,\, K+1 \leq m \leq 2K, \,\,m'=m-K, \nonumber
\end{eqnarray}
and $u^{[j]}_l$ and $v^{[j]}_l$ are the $l$th information symbols for the $j$th user in the BS 1 and BS 2, respectively.

The received signal at the $m$th time slot in phase 1 is
\begin{eqnarray} {\label{Eq3}}
\mathbf{Y}^{[i]}(m)=\underbrace{\sum^{K}_{j=1}\mathbf{H}^{[j,1]}_i(m)\mathbf{X}^{[j,1]}(m)}_{\underline{\triangle}\,\,\mathbf{L}^{[i]}_m(u^{[1]}_{m+1},u^{[2]}_{m+1},\cdots,u^{[K]}_{m})} + \mathbf{Z}^{[i]}(m),
\end{eqnarray}
and the signal in phase 2 is also described by
\begin{eqnarray}
\mathbf{Y}^{[i]}(m)=\underbrace{\sum^{K}_{j=1}\mathbf{H}^{[j,2]}_i(m)\mathbf{X}^{[j,2]}(m)}_{\underline{\triangle}\,\,\mathbf{L}^{[i]}_m(v^{[1]}_{m'+1},v^{[2]}_{m'+1},\cdots,v^{[K]}_{m'})} + \mathbf{Z}^{[i]}(m).
\end{eqnarray}

By the end of phase 1, BS 1 has $K^2$ equations in terms of $K(K+1)$ desired symbols, and thereby it needs $K$ extra linear independent equations to be able to resolve its desired symbols. Note that BS 2 also has $K^2$ (overheard) equations that contain no information for BS 2, but rather for BS 1. Thus, any $K$ linearly independent set of them can serve as the extra equations desired for BS 1. Similarly, the overheard equations saved by BS 1 during phase 2 can be extra equations for BS 2, since BS 2 has $K^2$ equations only while the total number of transmitted symbols in the phase is $K(K+1)$. In order to apply retrospective IA, we need to purify the overheard equations by proper linear combinations of them so that $K$ new linear equations per cell are solely in terms of information symbols of one transmitter as follows:\\
for $1 \leq m \leq K$,
 \begin{eqnarray}
\widehat{L}^{[2]}_m(u^{[m]}_{m},u^{[m]}_{m+1})=\mathbf{u}^{[1]\dag}_m\mathbf{L}^{[2]}_m(u^{[1]}_{m+1},u^{[2]}_{m+1},\cdots,u^{[K]}_{m}),
\end{eqnarray}
for $K+1 \leq m \leq 2K$,
 \begin{eqnarray}
\widehat{L}^{[1]}_m(v^{[m']}_{m'},v^{[m']}_{m'+1})=\mathbf{u}^{[2]\dag}_m\mathbf{L}^{[1]}_m(v^{[1]}_{m'+1},v^{[2]}_{m'+1},\cdots,v^{[K]}_{m'}).
\end{eqnarray}
where $\mathbf{u}^{[1]\dag}_m\mathbf{H}^{[\bar{m},1]}_2(m)\mathbf{X}^{[\bar{m},1]}(m) = 0$ for $1\leq m \leq K$, $\forall \bar{m}\in\{1,2,\cdots,K\},\bar{m}\neq m$ and $ \mathbf{u}^{[2]\dag}_m\mathbf{H}^{[\bar{m}',2]}_1(m')\mathbf{X}^{[\bar{m}',2]}(m') = 0$ for $K+1\leq m \leq 2K$, $\forall \bar{m}'\in\{1,2,\cdots,K\},\bar{m}'\neq m'$.
Note that the $2\times1$ vector $\mathbf{X}^{[\bar{m},1]}(m)$ has only one non-zero entry so that
$\mathbf{H}^{[\bar{m},1]}_2(m)\mathbf{X}^{[\bar{m},1]}(m)$ is effectively a $K \times 1$ matrix,
which indicates that $\mathbf{u}^{[1]\dag}_m$ always exists.
Without the purified process, each transmitter cannot locally generate the linear equations to deliver with delayed and local CSIT.

Thus, the goal of phase 3 is to swap these purified overheard equations for both receivers to resolve their intended symbols in an efficient manner.
With the help of the delayed local CSIT, each transmitter reconstructs the following transmit signals based on their own past transmitted symbols in phase 3;
\begin{eqnarray}\small
\bullet&\mathrm{Phase}\,\,3:& \mathbf{X}^{[j,1]}(2K+1) = \left[
                                                         \begin{array}{c}
                                                           \widehat{L}^{[2]}_j(u^{[j]}_{j},u^{[j]}_{j+1}) \\
                                                           0 \\
                                                         \end{array}
                                                       \right],\nonumber\\
                           &&\mathbf{X}^{[j,2]}(2K+1) = \left[
                             \begin{array}{c}
                               \widehat{L}^{[1]}_{j'}(v^{[j]}_{j},v^{[j]}_{j+1}) \nonumber\\
                               0 \\
                             \end{array}
                           \right],\nonumber\\
                           &&\mathrm{where}\,\, j'=j+K.
\end{eqnarray}

The received signal at each BS in phase 3 is described as:
\begin{eqnarray}
\mathbf{Y}^{[1]}(2K+1)={\sum^{K}_{j=1}\mathbf{h}^{[j,1]}_{1,1}(2K+1)\widehat{L}^{[2]}_j(u^{[j]}_{j},u^{[j]}_{j+1})}\hspace{3mm} \nonumber\\\hspace{-10mm}+
\underbrace{\sum^{K}_{j=1}\mathbf{h}^{[j,2]}_{1,1}(2K+1)\widehat{L}^{[1]}_{j'}(v^{[j]}_{j},v^{[j]}_{j+1})}_{\mathrm{inter-cell\,\, interference}}
+\mathbf{Z}^{[1]}(2K+1),\hspace{-3mm}\\
\mathbf{Y}^{[2]}(2K+1)=\underbrace{\sum^{K}_{j=1}\mathbf{h}^{[j,1]}_{2,1}(2K+1)\widehat{L}^{[2]}_j(u^{[j]}_{j},u^{[j]}_{j+1})}_{\mathrm{inter-cell\,\, interference}} \hspace{3mm}\nonumber\\+
{\sum^{K}_{j=1}\mathbf{h}^{[j,2]}_{2,1}(2K+1)\widehat{L}^{[1]}_{j'}(v^{[k]}_{j},v^{[j]}_{j+1})} +
\mathbf{Z}^{[2]}(2K+1),\hspace{-3mm}
\end{eqnarray}


Note that it is possible to eliminate inter-cell interference terms that come from users in the adjacent cell using the previously known purified overheard equations during phase 1 and phase 2 as side information as follows:
\begin{eqnarray}
&&{\mathbf{Y}^{[1]}(2K+1) - \sum^{K}_{j=1}\mathbf{h}^{[j,2]}_{1,1}(2K+1)\widehat{L}^{[1]}_{j'}(v^{[j]}_{j},v^{[j]}_{j+1})}\nonumber\\
 &&=  \left[
     \begin{array}{ccc}
       \mathbf{h}^{[1,1]}_{1,1}(2K+1)\,\,\,\cdots\,\,\,\mathbf{h}^{[K,1]}_{1,1}(2K+1)  \nonumber\\
     \end{array}
   \right]\\
&&\times\left[
  \begin{array}{c}
    \widehat{L}^{[2]}_1(u^{[1]}_{1},u^{[1]}_{2})  \\
    \vdots \\
    \widehat{L}^{[2]}_K(u^{[K]}_{K},u^{[K]}_{K+1}) \\
  \end{array}
\right]+\mathbf{{Z}}^{[1]}(2K+1).
\end{eqnarray}

Since all the elements of the channel matrix $\left[
     \begin{array}{ccc}
       \mathbf{h}^{[1,1]}_{1,1}(2K+1)\,\,\,\cdots\,\,\,\mathbf{h}^{[K,1]}_{1,1}(2K+1)  \nonumber\\
     \end{array}
   \right]$
are i.i.d., and the size of it is $K\times K$, the rank of the matrix becomes $K$ with probability one. Therefore, BS 1 simply can resolve $K$ unknown purified equations, observed and generated by the adjacent BS (i.e., BS 2) using simple a zero-forcing decoder. With the assistance of additionally resolved $K$ equations of desired symbols during phase 3, BS 1 can finally decode all desired $K(K+1)$ information symbols.

In the same argument, for the BS 2, they can also successfully resolve $K(K+1)$ data symbols by the end of phase 3. As a result, we can show that $\frac{2K(K+1)}{2K+1}$ sum-DoF can be achieved in total.

\section{Analysis of sum-DoF Gain and Optimality}
\label{sec_con}
\subsection{Sum-DoF Gain from Outdated CSIT}
To examine our achievable sum-DoF with delayed and local channel feedback, we first characterize the optimal sum-DoF under no CSIT assumption for $(K,2,K)$ MIMO-IMAC, and compare those two sum-DoF results. The sum-DoF outer bound in this channel is obtained by allowing perfect cooperation among $K$ users in each cell. If we assume perfect cooperation between the $K$ users, the $(K,2,K)$ MIMO-IMAC is converted into a two-user $(2K,K)$ MIMO-IC. Since cooperation does not hurt the capacity, the DoF region with no CSIT for the $(K,2,K)$ MIMO-IMAC is bounded as \cite{Vaze}-\cite{Huang}:
\begin{eqnarray}
\sum_{k=1}^K d^{[k,1]} \leq K,\,\,\,\,\,\, \sum_{k=1}^K d^{[k,2]} \leq K,\\
\sum_{k=1}^K d^{[k,1]} +\sum_{k=1}^K d^{[k,2]} \leq K.
\end{eqnarray}
Using the DoF outer bound region, one can prove that the optimal sum-DoF of the $(K,2,K)$ MIMO-IMAC with no CSIT, denoted by $\mathrm{DoF}^{\mathrm{\textit{No-CSIT}}}_{\mathrm{sum}}$, is equal to $K$. This is because a zero-forcing decoder at the receiver can achieve $K$ sum-DoF with no CSIT in $(K,2,K)$ MIMO-IMAC, thereby the achievable sum-DoF is tight in this setting.

Leveraging the results in Section III, we can compute the sum-DoF gain from the delayed CSIT over no CSIT:
\begin{eqnarray}
\mathrm{DoF_{sum}}&=&\frac{2K(K+1)}{2K+1},\\
&=&\mathrm{DoF}^{\mathrm{\textit{No-CSIT}}}_{\mathrm{sum}}\left(1+\underbrace{\frac{1}{2K+1}}_{\mathrm{growth\,\,\, factor}}\right),
\end{eqnarray}
where $\frac{1}{2K+1}$ represents the growth factor by taking advantage of the delayed CSI feedback rather than ignoring it.

\subsection{A Sum-DoF Outer Bound}
In this subsection, we derive a new sum-DoF outer bound by using \textit{Rank-Ratio Inequality} \cite{Avestimehr2} that can be applied for any arbitrary network where a receiver decodes its desired message in the presence of two interferers with delayed CSIT. In particular, we focus here on the $(M,M,2)$ MIMO-IMAC scenario in which there exist only two inter-cell interference signals so that the inequality holds in this case.

Now consider the decoding for user $[k,l]$ at BS $l$. The corresponding interference subspace at BS $l$ will be
\begin{eqnarray}
\boldsymbol{\mathcal{I}}_{[k,l]} = \mathrm{span}\left(\mathbf{H}^{[\bar{k},l]}_l \mathbf{V}^{[\bar{k},l]}\right)\hspace{35mm} \nonumber\\ \cup \mathrm{span}\left(\mathbf{H}^{[k,\bar{l}]}_l \mathbf{V}^{[k,\bar{l}]}\right)  \cup \mathrm{span}\left(\mathbf{H}^{[\bar{k},\bar{l}]}_l \mathbf{V}^{[\bar{k},\bar{l}]}\right),
\end{eqnarray}
where $\mathrm{span}(\cdot)$ of a matrix is the space spanned by its columns; $k,\bar{k},l,\bar{l}\in\{1,2\}$, and $k\neq \bar{k}$, $l\neq \bar{l}$. The decodability constraint for user $[k,l]$ at BS $l$ can be written as
\begin{eqnarray}
\mathrm{dim}\left(
\mathrm{Proj}_{\boldsymbol{\mathcal{I}}_{[k,l]}^c}\mathrm{span}\left(\mathbf{H}^{[k,l]}_l \mathbf{V}^{[k,l]} \right)
\right)
=\mathrm{dim}\left(
 \mathbf{V}^{[k,l]} \right)
=d^{[k,l]}, \nonumber
\end{eqnarray}
where $\boldsymbol{\mathcal{I}}_{[k,l]}^c$ denotes the subspace orthogonal to $\boldsymbol{\mathcal{I}}_{[k,l]}$; $\mathrm{Proj}_{\mathbf{B}^c}\mathrm{span}(\mathbf{A})$ is the orthogonal projection of column span of $\mathbf{A}$ on the orthogonal complement of the column span of $\mathbf{B}$.
Note that $\mathrm{dim}\left(\mathrm{span}\left(\mathbf{H}^{[k,l]}_l \mathbf{V}^{[k,l]} \right)\right) = \mathrm{dim}\left(\mathrm{span}\left(\mathbf{V}^{[k,l]} \right)\right)$ due to the continuous distribution of $\mathbf{H}^{[k,l]}_l$. By \textit{Lemma 3} in \cite{Avestimehr2}, the decodability condition can be rewritten as
{\small
\begin{eqnarray}
\label{Eq4}
\mathrm{rk}\left[
\mathbf{H}^{[{k},l]}_l \mathbf{V}^{[{k},l]}\right] +
\mathrm{rk}\left[
\mathbf{H}^{[\bar{k},l]}_l \mathbf{V}^{[\bar{k},l]}, \,\,\, \mathbf{H}^{[{k},\bar{l}]}_l \mathbf{V}^{[{k},\bar{l}]}, \,\,\,\mathbf{H}^{[\bar{k},\bar{l}]}_l \mathbf{V}^{[\bar{k},\bar{l}]}
\right]\nonumber\\
=\mathrm{rk}\left[
\mathbf{H}^{[{k},l]}_l \mathbf{V}^{[{k},l]}, \,\,\, \mathbf{H}^{[\bar{k},l]}_l \mathbf{V}^{[\bar{k},l]}, \,\,\, \mathbf{H}^{[{k},\bar{l}]}_l \mathbf{V}^{[{k},\bar{l}]}, \,\,\,\mathbf{H}^{[\bar{k},\bar{l}]}_l \mathbf{V}^{[\bar{k},\bar{l}]}
\right], \hspace{-0.8mm}
\end{eqnarray}}
where $\mathrm{rk}[\cdot]$ denotes the rank of a matrix.

\begin{figure*}[ht]
{\small
\begin{eqnarray}
\label{Eq5}
\mathrm{rk}\left[
\mathbf{H}^{[1,1]}_1 \mathbf{V}^{[1,1]}\right] +
\mathrm{rk}\left[
\mathbf{H}^{[2,1]}_1 \mathbf{V}^{[2,1]}\right] +
\mathrm{rk}\left[
\mathbf{H}^{[1,2]}_1 \mathbf{V}^{[1,2]}, \,\,\,\mathbf{H}^{[2,2]}_1 \mathbf{V}^{[2,2]}
\right]
=\mathrm{rk}\left[
\mathbf{H}^{[1,1]}_1 \mathbf{V}^{[1,1]}, \,\,\, \mathbf{H}^{[2,1]}_1 \mathbf{V}^{[2,1]}, \,\,\, \mathbf{H}^{[1,2]}_1 \mathbf{V}^{[1,2]}, \,\,\,\mathbf{H}^{[2,2]}_1 \mathbf{V}^{[2,2]}
\right],\\
\label{Eq6}
\mathrm{rk}\left[
\mathbf{H}^{[1,2]}_2 \mathbf{V}^{[1,2]}\right] +
\mathrm{rk}\left[
\mathbf{H}^{[2,2]}_2 \mathbf{V}^{[2,2]}\right] +
\mathrm{rk}\left[
\mathbf{H}^{[1,1]}_2 \mathbf{V}^{[1,1]}, \,\,\,\mathbf{H}^{[2,1]}_2 \mathbf{V}^{[2,1]}
\right]
=\mathrm{rk}\left[
\mathbf{H}^{[1,1]}_2 \mathbf{V}^{[1,1]}, \,\,\, \mathbf{H}^{[2,1]}_2 \mathbf{V}^{[2,1]}, \,\,\, \mathbf{H}^{[1,2]}_2 \mathbf{V}^{[1,2]}, \,\,\,\mathbf{H}^{[2,2]}_2 \mathbf{V}^{[2,2]}
\right],\hspace{0,5mm}
\end{eqnarray}}
\hrulefill
\end{figure*}

By \textit{Lemma 4} in \cite{Avestimehr2}, the equivalent decodability condition can be given by (\ref{Eq5})-(\ref{Eq6}) on the top of the next page. In addition, we have
{\small
\begin{eqnarray}
&&d^{[1,1]}+d^{[2,1]}+\frac{3}{2}\left(d^{[1,2]}+d^{[2,2]}\right)\\
&&=\mathrm{rk}\left[ \mathbf{V}^{[1,1]}\right] + \mathrm{rk}\left[ \mathbf{V}^{[2,1]}\right] +\frac{3}{2}\left(
\mathrm{rk}\left[ \mathbf{V}^{[1,2]}\right] + \mathrm{rk}\left[ \mathbf{V}^{[2,2]}\right]
\right),\nonumber\\
&&=\mathrm{rk}\left[\mathbf{H}^{[1,1]}_1 \mathbf{V}^{[1,1]}\right] + \mathrm{rk}\left[ \mathbf{H}^{[2,1]}_1\mathbf{V}^{[2,1]}\right] \nonumber\\
&&\hspace{22mm}+\frac{3}{2}\left(
\mathrm{rk}\left[\mathbf{H}^{[1,2]}_2 \mathbf{V}^{[1,2]}\right] + \mathrm{rk}\left[ \mathbf{H}^{[2,2]}_2\mathbf{V}^{[2,2]}\right]
\right),\nonumber
\end{eqnarray}}
which leads to
\begin{eqnarray}
\label{Eq11}
d^{[1,1]}+d^{[2,1]}+\frac{3}{2}\left(d^{[1,2]}+d^{[2,2]}\right)\leq \frac{3}{2}M,
\end{eqnarray}
by (\ref{Eq7})-(\ref{Eq9}) on the top of the page, where in (a) and (b), a basic property of the rank of a matrix, $\mathrm{rk}\left[ \mathbf{A}\,\,\,\, \mathbf{B} \right]\leq \mathrm{rk}\left[ \mathbf{A} \right] + \mathrm{rk}\left[\mathbf{B} \right]$ and \textit{Rank-Ratio Inequality} \cite{Avestimehr2} are used, respectively.
\begin{figure*}[ht]
{\small
\begin{eqnarray}
&&d^{[1,1]}+d^{[2,1]}+\frac{3}{2}\left(d^{[1,2]}+d^{[2,2]}\right)\nonumber\\
\label{Eq7}&&=\mathrm{rk}\left[
\mathbf{H}^{[1,1]}_1 \mathbf{V}^{[1,1]}, \,\,\, \mathbf{H}^{[2,1]}_1 \mathbf{V}^{[2,1]}, \,\,\, \mathbf{H}^{[1,2]}_1 \mathbf{V}^{[1,2]}, \,\,\,\mathbf{H}^{[2,2]}_1 \mathbf{V}^{[2,2]}
\right]-\mathrm{rk}\left[
\mathbf{H}^{[1,2]}_1 \mathbf{V}^{[1,2]}, \,\,\,\mathbf{H}^{[2,2]}_1 \mathbf{V}^{[2,2]}
\right]\nonumber\\ &&\,\,\,\,\,\,\,\,\,\,\,\,\,\,\,\,\,\,\,\,\,\,\,\,\,\,\,\,\,\,\,\,\,\,\,\,\,\,\,\,\,\,\,\,\,\,\,\,\,\,\,+ \frac{3}{2}
\left(
\mathrm{rk}\left[
\mathbf{H}^{[1,1]}_2 \mathbf{V}^{[1,1]}, \,\,\, \mathbf{H}^{[2,1]}_2 \mathbf{V}^{[2,1]}, \,\,\, \mathbf{H}^{[1,2]}_2 \mathbf{V}^{[1,2]}, \,\,\,\mathbf{H}^{[2,2]}_2 \mathbf{V}^{[2,2]}
\right]
-\mathrm{rk}\left[
\mathbf{H}^{[1,1]}_2 \mathbf{V}^{[1,1]}, \,\,\,\mathbf{H}^{[2,1]}_2 \mathbf{V}^{[2,1]}
\right]
\right),\\
\label{Eq8}&&\overset{(a)}{\leq} \mathrm{rk}\left[
\mathbf{H}^{[1,1]}_1 \mathbf{V}^{[1,1]}, \,\,\, \mathbf{H}^{[2,1]}_1 \mathbf{V}^{[2,1]} \right]\nonumber- \frac{3}{2}
\left(\mathrm{rk}\left[
\mathbf{H}^{[1,1]}_2 \mathbf{V}^{[1,1]}, \,\,\,\mathbf{H}^{[2,1]}_2 \mathbf{V}^{[2,1]}
\right]
\right)\\
&&\,\,\,\,\,\,\,\,\,\,\,\,\,\,\,\,\,\,\,\,\,\,\,\,\,\,\,\,\,\,\,\,\,\,\,\,\,\,\,\,\,\,\,\,\,\,\,\,\,\,\,+ \frac{3}{2}
\left(
\mathrm{rk}\left[
\mathbf{H}^{[1,1]}_2 \mathbf{V}^{[1,1]}, \,\,\, \mathbf{H}^{[2,1]}_2 \mathbf{V}^{[2,1]}, \,\,\, \mathbf{H}^{[1,2]}_2 \mathbf{V}^{[1,2]}, \,\,\,\mathbf{H}^{[2,2]}_2 \mathbf{V}^{[2,2]}
\right]
\right),\\
\label{Eq9}&&\overset{(b)}{\leq}\frac{3}{2}
\left(
\mathrm{rk}\left[
\mathbf{H}^{[1,1]}_2 \mathbf{V}^{[1,1]}, \,\,\, \mathbf{H}^{[2,1]}_2 \mathbf{V}^{[2,1]}, \,\,\, \mathbf{H}^{[1,2]}_2 \mathbf{V}^{[1,2]}, \,\,\,\mathbf{H}^{[2,2]}_2 \mathbf{V}^{[2,2]}
\right]
\right)\leq \frac{3}{2}M,
\end{eqnarray}}
\hrulefill
\end{figure*}
By symmetry, we are able to derive another bound,
\begin{eqnarray}
\label{Eq12}
\frac{3}{2}\left(d^{[1,1]}+d^{[2,1]}\right) +d^{[1,2]}+d^{[2,2]}\leq \frac{3}{2}M.
\end{eqnarray}
By combining the two bounds in (\ref{Eq11}) and (\ref{Eq12}), we can yield a new outer bound as
\begin{eqnarray}
d^{[1,1]}+d^{[2,1]} +d^{[1,2]}+d^{[2,2]}\leq \frac{6}{5}M.
\end{eqnarray}
This new outer bound coincides with the achievability for $(M,M,2)$ MIMO-IMAC as mentioned in \textit{Remark 1}, and the retrospective IA can achieve the optimal sum-DoF of $\frac{6}{5}M$.

\section{Conclusion}
\label{sec_con}
In this paper, a novel retrospective IA was proposed for the two-cell MIMO-IMAC with local and delayed CSIT. We have derived the achievable sum-DoF, and it was shown that the availability of delayed CSIT can strictly increase the sum-DoF over the no-CSIT case. This highlights the benefits of delayed and local CSIT for uplink cellular networks. Furthermore, we have proved that the retrospective IA can achieve the optimal sum-DoF for a special cases by providing a new outer bound.




\end{document}